\documentclass[twocolumn,showpacs,amsmath,amssymb,aps,pra,floatfix]{revtex4}
\usepackage{graphicx,graphics,times,color}
\usepackage{bm}
\usepackage{longtable}

\def\be{\begin{equation}}
\def\ee{\end{equation}}
\def\ba{\begin{eqnarray}}
\def\ea{\end{eqnarray}}
\def\la{\langle}
\def\ra{\rangle}

\begin{document}

\title{Entanglement Amplification in the Non-Perturbative Dynamics of Modular Quantum Systems}

\author{
A. Bayat$^{1}$, S. M. Giampaolo$^{2}$, F. Illuminati$^{2}$, and M. B. Plenio$^{1}$
}
\affiliation{
$^{1}$\mbox{Institut f\"{u}r Theoretische Physik, Albert-Einstein-Allee 11, Universit\"{a}t Ulm, D-89069 Ulm, Germany}
\\
$^{2}$\mbox{Dipartimento di Ingegneria Industriale, Universit\`a degli Studi di Salerno,~Via Ponte don Melillo, I-84084 Fisciano (SA), Italy}
}	

\begin{abstract}
We analyze the conditions for entanglement amplification between distant and not directly interacting quantum objects by their common
coupling to media with static modular structure and subject to a local (single-bond) quenched dynamics. We show that in the non-perturbative
regime of the dynamics the initial end-to-end entanglement is strongly amplified and, moreover, can be distributed efficiently between distant
objects. Due to its intrinsic local and non-perturbative nature the dynamics is fast and robust against thermal fluctuations, and its
control is undemanding. We show that the origin of entanglement amplification lies in the interference of the ground state and at most one of
the low-lying energy eigenstates. The scheme can be generalized to provide a fast and efficient router for generating entanglement between
simultaneous multiple users.
\end{abstract}

\date{March 12, 2013}


\maketitle

\section{Introduction}

Realizing a sizable and stable entanglement between distant qubits is a crucial stage in performing quantum information
and communication tasks~\cite{QI}. In the ground state of noncritical strongly correlated systems with short-range interactions the 2-points
correlation functions and thus the entanglement between spins vanish exponentially with the
distance~\cite{DecayofCorrelations}. To overcome this problem, a possibility is to mediate indirect interactions between distant qubits by a
suitable quantum many-body medium. For instance, one can exploit impurities weakly coupled to the ends of a spin chain to create a strong
long-distance entanglement between them~\cite{LDEcv1}. Indeed a relatively
large family of many-body systems allows for this possibility~\cite{LDEcv2,LDEsa2,LDEsa3}. However, for most of these
systems the energy gap is exponentially decreasing in the size of the system, so that even for short chains the mechanism becomes
thermally unstable: very small thermal fluctuations are sufficient to mix the ground state with higher energy eigenstates and suppress the
entanglement between the end impurities. One can introduce systems with interaction patterns such that the gap closes algebraically with the
distance~\cite{LDEcv2} at the price of letting the end-to-end entanglement become weakly
decreasing with the size of the system~\cite{LDEcv2,LDEsa2}. 
Even if all these results can be extended to higher dimensional systems~\cite{zipste} this appears to be the unavoidable limit
of a purely static, ground-state approach~\cite{Kuwahara}

On the other hand, it is well known that properly tailored time evolutions can propagate entanglement through many-body
systems~\cite{bose-review}. In particular, global~\cite{global-quench} or local quantum quenches~\cite{weak1,weak2,weak3,Bruderer,
bayat-dispatching} can create long-distance entanglement. Apart from certain perturbative proposals that suffer from a very slow
convergence~\cite{weak1,weak2,weak3,Bruderer}, the dynamical schemes in general do not require the weak end coupling
assumption~\cite{banchi-nonperturbative,bayat-dispatching,global-quench,bose-review} and hence they hold a greater resistance against
thermal instability but at the price of a much more elaborate control on the system. In conclusion, entangling distant
qubits can be realized either via static or dynamic approaches, however, either at the price of a strong thermal instability or an
excessively demanding control. 

In the present work we introduce a scheme that combines the modular-static and the quench-dynamic approaches to entanglement generation and
distribution among distant qubits. The proposed mixed scheme strongly reduces the drawbacks of both approaches and realizes a novel mechanism
of entanglement amplification starting from weakly entangled inputs. 
In our model, a quantum spin chain is split into two elementary bulk modules. Two end impurities are attached to each module with couplings
of arbitrary strength. This initial static configuration is then evolved through a sudden quench of the bond connecting the two modules.
We show that the generated end-to-end entanglement across the entire system is always larger than the initial entanglement at the ends of
each module and that this \emph{entanglement amplification} can always be achieved for all parameter ranges. As the length of each module is
half the total size of the system and the impurity couplings are non perturbative, the energy gap remains sizable and thus thermal stability is
assured. Moreover, by exploiting just a single bond quench
for inducing a nontrivial dynamics, the required control is minimal. This mechanism can demonstrate an entanglement router which unlike
previous proposals neither needs AC-control of the couplings~\cite{router-giampaolo-hanggi}
nor the presence of both ferromagnetic and anti-ferromagnetic couplings simultaneously~\cite{router-kay}.
Furthermore, the entanglement amplification mechanism finds a clear physical explanation in the quantum interference between only two
eigenvectors of the final Hamiltonian.

\begin{figure}
\centering
     \includegraphics[width=7cm,height=6cm,angle=0]{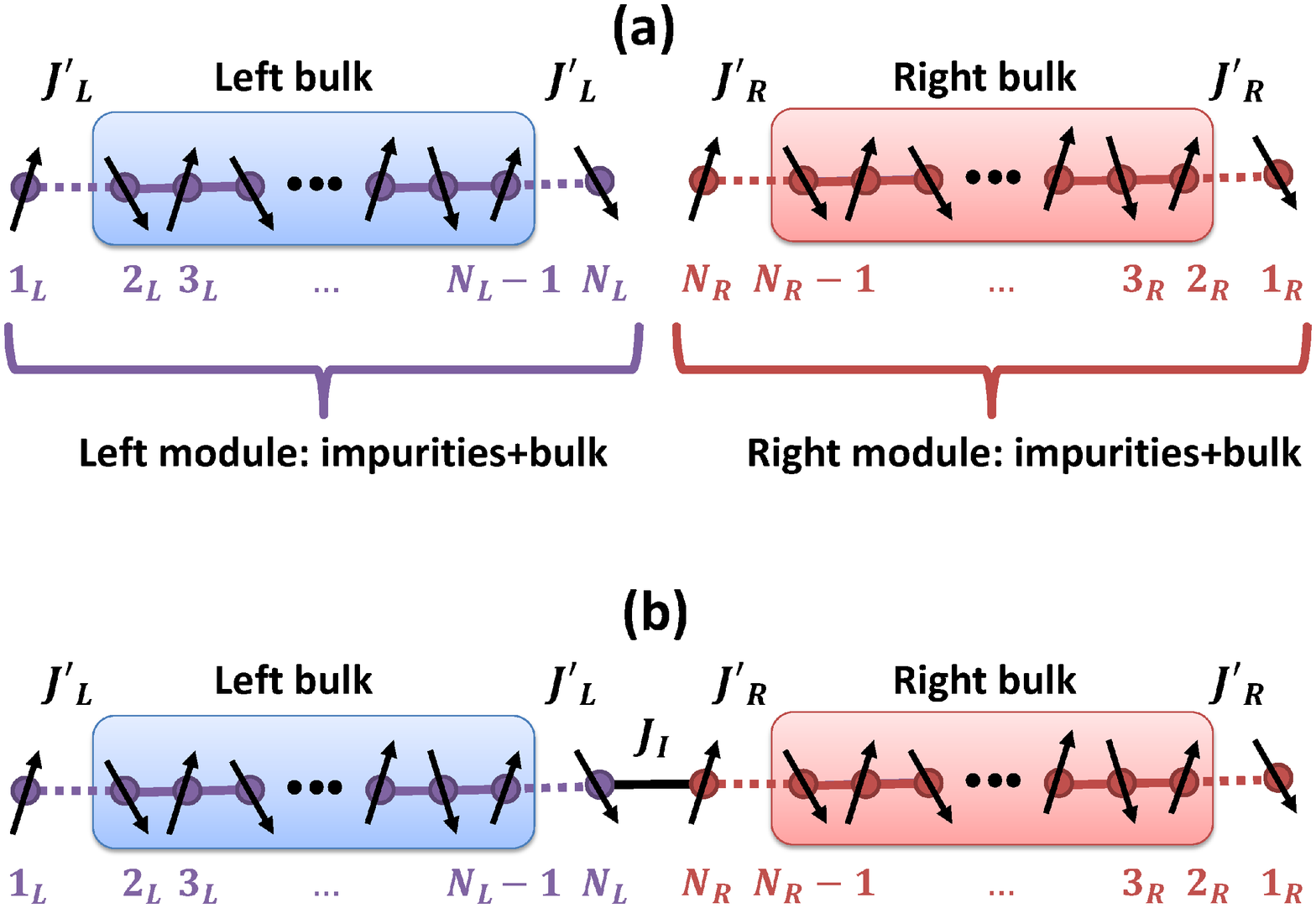}
     \caption{(Color online) (a) A schematic picture of two independent modules and their couplings to the respective end impurities.
     (b) Onset of a quenched interaction dynamics between the two modules by switching on instantaneously a strong bond $J_I$ between the
     impurities at the module-module boundary.}
      \label{fig1}
\end{figure}

\section{Introducing The Model}
We consider two chains of qubits, the {\it modules}, each constituted by a bulk of
spins coupled at both ends with two impurities. The Hamiltonian of each module reads:
\begin{eqnarray}
\label{Hxxz}
H_k &=& J'_kJ (h^{(k)}_{1,2}+h^{(k)}_{N_k-1,N_k})+J\sum_{i=2}^{N_k-2} h^{(k)}_{i,i+1} \; , \cr
h^{(k)}_{i,j} &=& X^{(k)}_i X^{(k)}_j+Y^{(k)}_i Y^{(k)}_j+\delta Z^{(k)}_i Z^{(k)}_j \; ,
\end{eqnarray}
where $k=\!L,R$ denotes the left or right module, respectively with $N_k \! = \! N_L \!, \!N_R$ spins,
$\{ \! X^{(k)}_i\!, \! Y^{(k)}_i\! ,\! Z^{(k)}_i \! \}$ are the Pauli spin operators at site $i$ in the $k$-th module,
$J\! > \! 0$ is the exchange spin-spin coupling strength, the dimensionless parameter $J'_k\!>\!0$ specifies the coupling to the end
impurities, and $\delta$ is the interaction anisotropy along the $z$ direction. A schematic picture
of this system is shown in Fig.~\ref{fig1}~(a) (note the mirror inversion of spin numbering in each module).
The spin-$1/2$ XXZ Hamiltonian Eq.~(\ref{Hxxz}) for each module has a rich zero-temperature quantum phase diagram; in particular, for
$-1 \! \leq  \! \delta \! \leq \! 1$, the system is in the so-called gapless XY antiferromagnetic phase that admits a
nondegenerate, highly entangled ground state. In the two fully isotropic limits one recovers the relevant cases of the XX Hamiltonian
($\delta\! =\! 0$) and the Heisenberg (XXX) Hamiltonian ($\delta\! =\! 1$).

\begin{figure}
\centering
\includegraphics[width=8cm,height=4cm,angle=0]{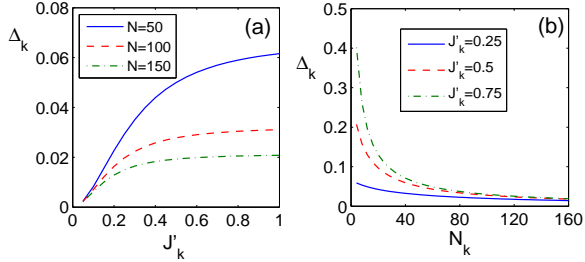}
\caption{(Color online) Energy gap $\Delta_k$ of a single XX module as a function of: (a) the coupling $J'_K$ to the end impurities for different
sizes $N_k$ of the module; (b) the size $N_k$ for different values of $J'_k$.}
\label{fig2}
\end{figure}

\section{Static end-to-end entanglement in modules}

The model in Eq.~(\ref{Hxxz}) exhibits long-distance entanglement in the ground state when the module has
\emph{even} length $N_k$ and $J'_k \! < \! J$ so that the spins in the bulk tend to entangle between themselves and consequently the two end
impurities are forced to get entangled~\cite{LDEcv2,LDEsa2}. From the ground state of the XXZ Hamiltonian $H_k$, one can compute the
initial reduced density matrix of the two impurities by tracing out the spins of the bulk. Due to the symmetries of the system, the reduced
state is diagonal in the Bell basis
\begin{equation}\label{rhoin}
 \rho_{1_k,N_k}=P^s_{k}|\psi^-\ra \la\psi^-|+ \sum_{\alpha=x,y,z} P^\alpha_{k} |B^\alpha\ra \la B^\alpha|,
 \end{equation}
where $|\psi^-\ra$ is the singlet state, $P^s_{k}$ is the singlet fraction, $|B^\alpha\ra$ (for $\alpha=x,y,z$) are the other Bell states
which can be obtained by applying the $\alpha$ Pauli operator on one part of the singlet, and the $P^\alpha_{k}$'s account for their
contributions. For a fixed length $N_k$, one can increase $P^s_{k}$ arbitrarily and create large entanglement between the two impurities by
decreasing $J'_k$. However, by decreasing $J'_k$ the energy gap of the system also decreases, at best algebraically~\cite{LDEcv2,LDEsa2};
when the thermal energy $k_BT$ becomes comparable with the gap, the state of the system becomes a mixed thermal state with no
entanglement between the end impurities. In Fig.~\ref{fig2}(a) the energy gap $\Delta_k$ is plotted as a function of $J'_k$ for a XX module of
length $N_k$ which clearly shows an exponential decay for small $J'_k$'s. Of course, the gap also decays with the size $N_k$ as shown in
Fig.~\ref{fig2}(b).

\section{Dynamical entanglement generation in a modular quantum spin chain}

Given the initial static situation with the two modules in their
respective ground states, we introduce a quench dynamics between them. The two even-sized modules of lengths
$N_L$ and $N_R$ are described via the Hamiltonians $H_L$ and $H_R$ introduced in Eq.~(\ref{Hxxz}).
The exchange coupling $J$ and anisotropic parameter $\delta$ are assumed to be identical in both
modules, while the couplings with the end impurities  are respectively $J'_L$ and $J'_R$. The initial ground state of a chain of total lengt
$N\!=\!N_L\!+\!N_R$ formed by the two noninteracting modules is obviously the tensor product of the ground states of the two subsystems:
$|\psi(0)\ra\!=\!|GS_L\ra \! \otimes\! |GS_R\ra$. At $t\!=\!0$ one switches on the interaction
between the two modules:
\begin{equation}\label{HI}
H_I=J_I J (X^{(L)}_{N_L} X^{(R)}_{N_R}+ Y^{(L)}_{N_L} Y^{(R)}_{N_R}+\delta Z^{(L)}_{N_L} Z^{(R)}_{N_R}) \; ,
\end{equation}
where the dimensionless parameter $J_I$ is the bond that couples the two modules. The total Hamiltonian of the system at $t\!\!>0$ becomes,
as schematicly represented in
Fig.~\ref{fig1}(b), $H_{T}\! =\! H_L\! +\! H_R \! + \! H_I$. The initial product state is not an eigenvector of $H_T$ and the system evolves
according to $|\psi(t)\ra\!=\!e^{-iH_{T}t}\!|\psi(0)\ra$. The reduced state of the two end impurities at time $t$ is
\begin{equation}\label{rhoout} \nonumber
\rho_{1_L,1_R}(t)=P^s_{out}(t)|\psi^-\ra \la\psi^-| + \sum_{\alpha=x,y,z} P^\alpha_{out}(t) |B^\alpha\ra \la B^\alpha|.
\end{equation}
We use relative entropy of entanglement~\cite{VedralP97} as an operational measure to quantify the entanglement content of the two ending
impurities at sites $1_L$ and $1_R$ defined by
\begin{equation}\label{Ent_t}
E_{1_L,1_R}(t)=1 - H\left( P^s_{out}(t) \right)
\end{equation}
where $H(x)\!=\!-x\!\log_2\! x\! -\! (\!1\!-\!x\!)\!\log_2\!(\!1\!-\!x\!)$.
This is of course related to the concurrence~\cite{Wootters} of the two impurities $C_{1_L,1_R}(t)$ via
$E_{1_L,1_R}(t)\!=\!1\! -\! H\left((C_{1_L,1_R}(t)\!+\!1)/2\right)$ and always give a lower value with respect to concurrence.


\subsection{Perturbative regime}

We first study the case of $J'_L$ and $J'_R$ both sufficiently small
so that the pairs of end impurities in each module are initially highly entangled (strong initial end-to-end entanglement in both modules).
In this situation, in each module the impurities are effectively decoupled from the rest of the system and resorting to the Schrieffer-Wolff
transformation~\cite{Bravyi-DiVincenzo-Loss} one obtains an effective interaction Hamiltonian for the impurities in each module
\begin{equation}\label{Heff}
H^k_{eff}=J^k_{eff} (X^k_{1_k}X^k_{N_k}+Y^k_{1_k} Y^k_{N_k}+\delta Z^k_{1_k}Z^k_{N_k}) \; ,
\end{equation}
where the effective coupling is linear in the energy gap: $J^k_{eff}\!=\!\Delta_k/{4}$ with $k\!=\!L,R$.
Fixing the inter-module interaction bond at $J_I\!=\!J^L_{eff}\!+\!J^R_{eff}$, as shown in Fig.~\ref{fig1}(b), one has
that $E_{1_L,1_R}(t)\!=\!1\!-\!H\left(\frac{5-3\cos(4J_It)}{8}\right)$. Therefore, at the optimal time $t_{opt}\!=\!\frac{\pi}{4J_I}$,
the maximal long-distance entanglement is established between the ending sites.
Compared to the static case with a single long module of length $N$ with the same entanglement between the end impurities $1_L$ and $1_R$,
the dynamical mechanism in the perturbative regime assures a larger thermal stability "per s\'e". If the entangling time $t_{opt}$ is
engineered so to be much smaller than the thermalization time, then the thermal effects are fully determined by the thermal initial state
associated to the energy gap of each module, which is always above and can be made much larger than the gap of the two combined in a single
one of size $N \!=\! N_L \!+\! N_R$. In the dynamic case one can exploit larger impurity couplings $J'_k$ for each module in comparison to the
static case with a single module of size $N$, thus increasing the gap, as illustrated in Fig.~\ref{fig2}(a). In this way thermal instability
is ameliorated but not fixed; moreover, the perturbative nature of the coupling $J_I$ implies a very slow dynamics.

\begin{figure}
\centering
\includegraphics[width=8cm,height=6cm,angle=0]{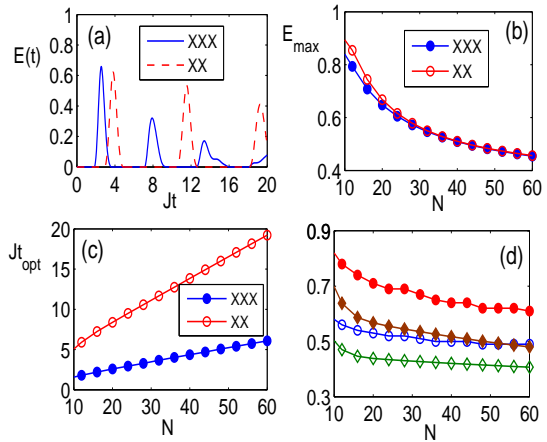}
\caption{\!(Color online)\! (a) End-to-end entanglement as a function of time for $N_L\!\!=\!\!N_R\!\!=\!\!10$,\! $J'\!\!=\!\!0.5$
and $J_I\!\!=\!\!0.65$.\! (b) $E_{max}$\! as a function of $N$.\! (c) Optimal time $t_{opt}$\! versus length $N$.\! (d) Optimal impurity
coupling $J'_{opt}$\! for XXX
(open blue circles) and  XX (green open diamonds) and optimal interaction coupling
$J_I^{opt}$\! for XXX (filled red circles) and
 XX (filled brown diamonds).}
\label{fig3}
 \end{figure}

\subsection{Non-perturbative regime: Entanglement amplification}

When the strength of the coupling to the impurities becomes comparable to the
interaction energies in the bulk, the initial end-to-end entanglement for each module is strongly suppressed, the reduction to the effective
Hamiltonians Eq.~(\ref{Heff}) is no longer justified, and $J_I$ cannot be determined analytically as in the perturbative regime. To proceed,
one can observe that in practice the affordable time for the entangling dynamics is ultimately bounded by the decoherence rates. Introducing
such upper bounds for the optimal entangling time, we define an optimization problem for the largest attainable end-to-end entanglement that
depends on two free parameters, i.e. the static coupling to the end impurities $J'\!=\!J'_L\!=\!J'_R$ and the quench-dynamic bond $J_I$
between the two modules.
In Fig.~\ref{fig3}(a) the end-to-end entanglement $E_{1_L\!,\!1_R}(t)$ is plotted as a function of time both for the Heisenberg and the XX
Hamiltonians. As reported in Fig.~\ref{fig3}(a), in both the two cases for such a choice of $J'$ and $J$, the time evolution after the quench
generates a large end-to-end entanglement, strongly amplified respect to the initial one in each module. The optimal entangling time $t_{opt}$ is
then the earliest time at which the end-to-end entanglement peaks, defining its maximum attainable value:
$E_{max}\!=\!E(t_{opt})$. In Fig.~\ref{fig3}(b) we plot $E_{max}$ as a function of the total chain's size $N$ when both $J'$ and $J_I$ are
tuned to their optimal values. One has that for long enough chains both the XX and Heisenberg Hamiltonians perform equally well. However,
if we look at Fig.~\ref{fig3}(c), where $t_{opt}$ is reported as a function of $N$, one can see that the Heisenberg chain generates a faster
dynamics with earlier peaks, an important added value in order to minimize the effects of decoherence. Finally, in Fig.~\ref{fig3}(d) we
report the optimal values $J'_{opt}$ and $J_I^{opt}$, respectively of the coupling to the end impurities and the interaction coupling in the
bulk, as functions of $N$. Remarkably, as
Fig.~\ref{fig3}(d) clearly shows, one finds that the optimal couplings decrease very slowly by increasing the size $N$, implying the onset
of a fast dynamics and the permanence of the non-perturbative regime even for very long chains. Furthermore, larger couplings in Heisenberg
chains in compare to XX ones results in a faster dynamics and provides higher energy gap and thus higher thermal stability.

\begin{figure}
\centering
     \includegraphics[width=8cm,height=4cm,angle=0]{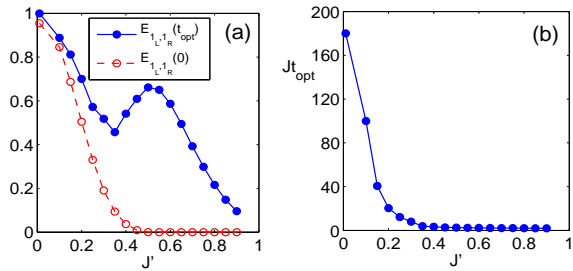}
     \caption{\!(Color online)\! (a) The initial, static end-to-end entanglement $E_{1_L,N_L}(0)$ (or $E_{1_R,N_R}(0)$) in each single module of
     length $N_L\! =\! N_R\! =\!10$ together with the maximal end-to-end dynamical entanglement $E_{1_L,1_R}(t_{opt}$
     for the entire two-module chain of length $N \!=\! N_L\! +\! N_R\!$, as functions of the dimensionless coupling $J'$ for $\delta\!=\!1$.
\!(b) The optimal time $Jt_{opt}$ (in the units of $\hbar$) at which entanglement peaks, as a function of $J'$. }
      \label{fig4}
 \end{figure}

To show how entanglement is actually amplified by the dynamical process we compare the initial static end-to-end entanglement
$E_{1_L,N_L}(0)$ in the ground state of a single module of length $N_L\! = \!N/2$ and the output dynamical end-to-end entanglement
$E_{1_L,1_R}(t_{opt})$ at the optimal entangling time across two interacting modules forming a chain of total length $N_L\! +\! N_R\! =\! N$.
Fig.~\ref{fig4}(a) shows that the initial ground-state end-to-end entanglement is {\it always} amplified for the entire range of couplings
$J'$ to the end impurities. A remarkable rebound of $E_{1_L,1_R}(t_{opt})$ occurs at a point of non-analyticity that separates the
perturbative from the non-perturbative regime, corresponding to the onset of enhanced dynamical amplification against pure decay. In order
to characterize and understand these two different regimes and the transition between them we have studied the entangling time $t_{opt}$ as a
function of the coupling $J'$ to the end impurities. Fig.~\ref{fig4} (b) illustrates
that in the perturbative regime $t_{opt}$ increases exponentially with decreasing $J'$, while in the non-perturbative regime it remains
essentially flat at very small values, guaranteeing, for appropriately chosen values of $J'$, that maximal amplification occurs well in
advance of the effects of decoherence.

\section{Origin of amplification: Excitation spectrum and quantum interference}

\begin{figure}[t]
\centering
    \includegraphics[width=8cm,height=6cm,angle=0]{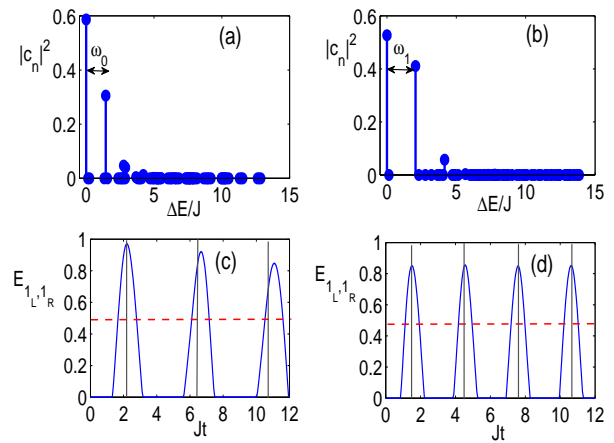}
    \caption{\!(Color online) Upper panels: $|c_n^2|$ versus $\Delta\! E/J\!=\!(E_n\!-\!E_0)/J$ for all eigenvectors of
    $H_T$ for two chains with $N_L\!=\!N_R=4$,
    $J'_L\!=\!J'_R\!=\!0.5\! J$,\! $J_I\!=\!0.75\! J$ where $J$ is the bulk amplitude for (a) $\delta\!=\!0$; (b) $\delta\!=\!1$.
    Lower panels:
    $E_{1_L,1_R}$ vs. time for the two chains with parameters given above and (c) $\delta\!=\!0$; (d) $\delta\!=\!1$.
    The red dashed lines stand for the
    initial end to end entanglement in each single chain. The vertical gray lines signal the times at which the phase
    $\phi(t)\!=\!\exp(-i \omega_\delta\! t)\!=\!-1$.}
     \label{new_F5}
\end{figure}

To understand the physical mechanism underlying entanglement amplification in the non-perturbative dynamics of modular many-body systems we
must take notice of the fact that \!$|\psi(t)\ra\!=\!\sum_{n=1}^{2^N}\!c_ne^{-iE_nt}|E_n\ra$, where, $E_n$\! ($|E_n\ra$) is the $n$-th
eigenvalue (eigenvector) of $H_T$ and $c_n\!=\!\la E_n\!| \!\psi(0)\ra$. When the values of the Hamiltonian parameters $J'$ and $J_I$ are
far from their optimal values, many of $c_n$'s are significantly different from zero. At values close to the optimal set the situation becomes
radically different, as reported in Figs.~\ref{new_F5}(a) and (b), where the squared amplitudes $|c_n|^2$ are plotted as
functions of the energy difference between the corresponding $n$-th eigenstate and the ground state of $H_T$, respectively for the XX
($\delta=0$, Fig.~\ref{new_F5}(a)) and the Heisenberg Hamiltonian ($\delta\!=\!1$, Fig.~\ref{new_F5}(a)). In both cases the initial state is
essentially projected onto only two eigenstates of $H_T$, the ground state and just one of the first few low-lying excited states. Therefore
the evolution of the initial state at time $t$ under the action of $H_T$ is essentially due to the relative phase factor
$\phi(t)\!=\!\exp(-i \omega_{\delta} t)$, where $\omega_{\delta}$ is the the difference of the eigenvalues of the two eigenstates
as indicated in Figs.~\ref{new_F5}(a) and (b). As shown in Figs.~\ref{new_F5}(c) and(d), the maximum of the entanglement amplification is
reached when $\phi(t) \simeq -1$. Therefore the dynamic amplification of long-distance entanglement is essentially due to a constructive
interference between the two eigenstate while the fact that the maximum is reached around and not exactly at $\phi(t)=-1$ accounts for the
projections on the remaining part of the spectrum: although strongly suppressed, they are not exactly vanishing. As we move to values of the
Hamiltonian parameters away from the optimal ones and/or the dimension of the two modules is increased, the relative weight of the other
eigenstates becomes more relevant reducing the maximum value reached by $E_{1_L,1_R}$.

It is worth mentioning that in perturbative dynamics we also see the interference between two eigenstates of the
Hamiltonian~\cite{weak1,weak2,weak3,Bruderer}, however, those eigenvectors have \emph{localized} excitations on edges and have the same
structure in the bulk. Due to this minimal difference, i.e. only at the edges where the sender and receiver are located, they are separated by a
vanishing energy in the spectrum of the system which results in a very slow dynamics. In contrast, the relevant two eigenstates in our dynamics
have completely distinct global structures which also make their energy separations sensibly large and result in our fast dynamics.

\section{Imperfections}

There are a few imperfection effects which may affect the ideal evolution presented above. Here we consider two main effects: random couplings
and thermal fluctuations.

\subsection{Random Couplings}

The first effect that we consider here is the randomness in exchange couplings. We assume that $J\rightarrow J(1+\epsilon)$ where $\epsilon$ is a
dimensionless random number with uniform distribution over $[-\lambda,+\lambda]$. We consider two modules tuned to their optimal values of
$J_I$ and $J'$ and a random noise is affecting all the couplings of the system. Due to the random change in the couplings the optimal time at
which entanglement peaks may also change. However, determination of this change is only possible if the real values of the random couplings are
known. If the real values are not known we have to set the time equal to its value for the clean system. In TABLE I we give the average values of
entanglement over 50 different realization of the $XX$ system at its peak (i.e. $<E_{max}>$) and the average of entanglement at the optimal time
$t_{opt}$, determined for the clean system (i.e. $<E(t_{opt})>$), for different values of $\lambda$. As it is clear from the TABLE I the
attainable entanglement deteriorates by increasing $\lambda$ while the average time at which entanglement peaks changes very slowly.
In TABLE 2, the same data for the Heisenberg Hamiltonian is shown. Comparing the two table is possible to see how the $XX$ systems results
much more noise resistant than the Heisenberg one. Such fact that is associated to the inner structure of the spectrum of the eigenstate of the
Hamiltonian makes the XX model more interesting when in our experimental realization the main source of noise is associated to a lack of precision
in the interactions between spins.

\begin{table}
\begin{centering}
\begin{tabular}{|c|c|c|c|c|c|}
  \hline
  $\lambda$        & 0.0000  & 0.0500  & 0.1000  & 0.1500  & 0.2000    \\
  \hline
  $<E_{max}>$      & 0.7442  & 0.7256  & 0.6812  & 0.6029  & 0.5391   \\
  \hline
  $<E(t_{opt})>$   & 0.7442  & 0.7224  & 0.6727  & 0.5839  & 0.5005   \\
  \hline
  $<Jt_{peak}>$     & 7.2446  & 7.2427  & 7.2557  & 7.2718  & 7.3007   \\
  \hline
\end{tabular}
\caption{ The effect of random couplings on $XX$ chain of length $N_L=N_R=8$. }
\par\end{centering}
\centering{}\label{table_1}
\end{table}

\begin{table}
\begin{centering}
\begin{tabular}{|c|c|c|c|c|c|}
  \hline
  $\lambda$        & 0.0000   & 0.0500  & 0.1000  & 0.1500  & 0.2000    \\
  \hline
  $<E_{max}>$        & 0.7442   & 0.6710  & 0.5941  & 0.5028  &  0.4106   \\
  \hline
  $<E(t_{opt})>$   & 0.7442   & 0.6380  &  0.5549 & 0.4659  &  0.3820  \\
  \hline
  $<Jt_{peak}>$     & 2.2000   & 2.1200  & 2.1000  &  2.0900 &  2.0900   \\
  \hline
\end{tabular}
\caption{ The effect of random couplings on a $XXX$ chain of length $N_L=N_R=8$. }
\par\end{centering}
\centering{}\label{table_2}
\end{table}

\subsection{Thermal Fluctuations}

\begin{figure}[t]
\centering
     \includegraphics[width=8cm,height=4cm,angle=0]{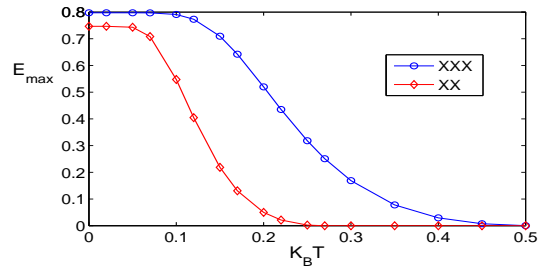}
     \caption{\!(Color online)\! The maximum attainable entanglement versus temperature $K_BT$ in a system of $N_L=N_R=6$ for the both
     $XX$ and $XXX$ chains.}
      \label{fig6}
 \end{figure}

The main issue of exploiting dynamics is to overcome the thermal instability of the static entanglement. In this section we analyze the
destructive effect of thermal fluctuations in our system. We assume that the modules cannot be cooled to their ground state and they are
initially in a thermal state  as $\rho(0)=e^{-(H_L+H_R)/K_BT}$, where $T$ is temperature and $K$ is Boltzmann constant. The evolution of the system
is then given by
\begin{equation}\label{evolution_rhot}
    \rho(t)=e^{-iH_Tt}\rho(0)e^{+iH_Tt}.
\end{equation}
From $\rho(t)$ one can get the density matrix of the boundary spins by tracing out the rest and compute its entanglement as before. In
Fig.~\ref{fig6} the $E_{max}$ is plotted as a function of temperature for both $XX$ and $XXX$ chains. As these figures clearly show there is a
plateau in small temperatures which its width is proportional to the gap of the system. The wider plateau observed for the Heisenberg chain in
compare to the $XX$ one is in fact due to its larger energy gap and results in a better thermal stability for the Heisenberg interaction.
Therefore if in our experimental facility we expect that if the greatest source of noise is due to the thermal effects, than we must
choose to realize the entanglement amplifier using Heisenberg chains.
The time at which entanglement peaks hardly changes with temperature which is in agreement with \cite{bayat-thermal}.

\section{Conclusions \& Outlook: Entanglement router}

By considering many-body media with static structures of bulk modules and by engineering suitably quenched local interaction dynamics
between different modules, we have introduced a method for the generation and amplification of entanglement between distant and non-interacting
quantum objects. The method is intrinsically non-perturbative and does not require sophisticated controls of the system dynamics. Indeed, we
showed that the end-to-end entanglement initially present in the ground state of a modular many-body system is amplified by the quenched dynamics
whenever all the values of the Hamiltonian parameters are far from the perturbative regime. The ensuing entangling dynamics is fast and robust
against thermal fluctuations as well as against the presence of noise in the amplitude interactions.
The occurrence of such dynamical mechanism of entanglement amplification finds a simple and beautiful explanation in the constructive
interference of only two energy eigenstates of the driving Hamiltonian. This opens potentially many new perspective for studying entanglement
generation, distribution, and manipulation across arbitrary distances in a many-body systems.
Our minimal control strategy applies immediately to the demonstration of an entanglement router capable of distributing entanglement
simultaneously between multiple users. Indeed, consider a setup in which every user controls one impurity in a
module which extends from its position to a common dispatching center where the other impurities are controlled. In the dispatching center,
one can switch on the interaction between any pair of impurities of different modules and induce the entangling dynamics in those
particular chains, thus creating entanglement directly between the two users. The main advantage of such an entanglement router is the minimal
control needed (a single bond quench) that, at variance with previous router proposals, neither requires AC-control of the
couplings~\cite{router-giampaolo-hanggi} nor the introduction of many different couplings~\cite{router-kay}. All this advantages make the
present approach suitable of a relatively easy experimental realizations based on different experimental devices as the doped coupling optical
cavities~\cite{LDEsa2,coupled-cavities} or the trapped ions~\cite{trapped-ions}

\section*{ Acknowledgements:} AB and MBP acknowledge financial support from the EU STREP project PICC and the Alexander von Humboldt foundation.
SMG and FI acknowledge financial support from the EU STREP Project iQIT, Grant Agreement No. 270843.

\end{document}